\documentclass[nobibnotes,preprintnumbers,aps,prl,superscriptaddress,showpacs,twocolumn,twoside,sort&compress]{revtex4}				 %
\usepackage{graphicx}																												 %
\usepackage{titlesec}
\usepackage{amsmath}																											     %
\usepackage{fancyheadings}																							 				 %
\begin{document}																													 %
\preprint{{Vol.XXX (201X) ~~~~~~~~~~~~~~~~~~~~~~~~~~~~~~~~~~~~~~~~~~~~~~~~~~~~ {\it CSMAG`16}										  ~~~~~~~~~~~~~~~~~~~~~~~~~~~~~~~~~~~~~~~~~~~~~~~~~~~~~~~~~~~~ No.X~~~~}}																 %
\vspace*{-0.3cm}																													 %
\preprint{\rule{\textwidth}{0.5pt}}																											 \vspace*{0.3cm}																														 %

\title{Effect of the Canting of Local Anisotropy Axes on Ground-State Properties of a Ferrimagnetic Chain with Regularly Alternating Ising and Heisenberg Spins}
\author{J. Torrico}
\thanks{corresponding author; e-mail: jordanatorrico@gmail.com}
\author{M.L. Lyra}
\affiliation{Instituto de F\'isica, Universidade Federal de Alagoas, 57072-970 Maceio, AL, Brazil}
\author{O. Rojas}
\author{S.M. de Souza}
\affiliation{Departamento de F\'isica, Universidade Federal de Lavras, 37200-000, Lavras-MG}
\author{J. Stre\v{c}ka}
\affiliation{Institute of Physics, Faculty of Science, P. J. \v{S}af\'arik University, Park Angelinum 9, 040 01 Ko\v{s}ice, Slovakia}

\begin{abstract}
The effect of the canting of local anisotropy axes on the ground-state phase diagram and magnetization of a ferrimagnetic chain with regularly alternating Ising and Heisenberg spins is exactly examined in an arbitrarily oriented magnetic field. It is shown that individual contributions of Ising and Heisenberg spins to the total magnetization basically depend on the spatial orientation of the magnetic field and the canting angle between two different local anisotropy axes of the Ising spins.
\end{abstract}

\pacs{75.10.Pq ; 75.10.Kt ; 75.30.Kz ; 75.40.Cx ; 75.60.Ej}
\maketitle

\section{Introduction}
In spite of a certain over-simplification,  a few exactly solved Ising-Heisenberg models capture essential magnetic features of some real polymeric coordination compounds as for instance Cu(3-Clpy)$_2$(N$_3$)$_2$ \cite{str05}, [(CuL)$_2$Dy][Mo(CN)$_8$] \cite{heu10,bel14} and [Fe(H$_2$O)(L)][Nb(CN)$_8$][Fe(L)] \cite{sah12}. The rigorous solutions for the Ising-Heisenberg models thus afford an excellent playground for experimental testing of a lot of intriguing magnetic properties such as quantized magnetization plateaus, anomalous thermodynamics, enhanced magnetocaloric effect, etc. \cite{str05,heu10,bel14,sah12}

Recently, it has been verified that the bimetallic coordination polymer Dy(NO$_3$)(DMSO)$_2$Cu(opba)(DMSO)$_2$ (to be further abbreviated as DyCu) can be satisfactorily described by the spin-1/2 Ising-Heisenberg chain with regularly alternating Ising and Heisenberg spins, which capture the magnetic behavior of Dy$^{3+}$ and Cu$^{2+}$ magnetic ions, respectively \cite{str12}. However, a closer inspection of available structural data reveals two crystallographically inequivalent orientantions of coordination polyhedra of Dy$^{3+}$ magnetic ions, which regularly alternate along the DyCu chain \cite{cal08}. Motivated by this fact, we will investigate in the present work the effect of the canting between two different local anisotropy axes on the ground-state properties of the spin-1/2 Ising-Heisenberg chain with regularly alternating Ising and Heisenberg spins in an arbitrarily oriented magnetic field.

\section{Model and its Hamiltonian}

\begin{figure}[htbp]
\includegraphics[scale=0.55]{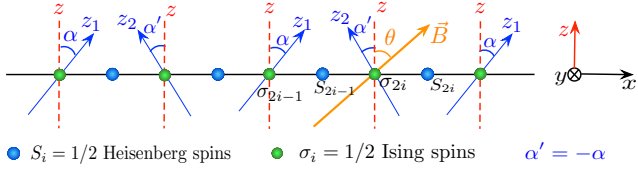} \caption{\label{fig:1} (Color online) Schematic representation of a spin chain with regularly alternating Ising and Heisenberg spins. The angle $\alpha$ (-$\alpha$) determines the canting of the local anisotropy axis $z_1$ ($z_2$) from the global frame $z$-axis for odd (even) Ising spins so that $2 \alpha$ is the canting angle between two coplanar anisotropy axes. The angle $\theta$ determines the tilting of the magnetic field from the global frame $z$-axis.}
\end{figure}

Let us introduce the spin-1/2 Ising-Heisenberg chain schematically illustrated in Fig.~\ref{fig:1}, in which the Ising spins with two different local anisotropy axes $z_1$ and $z_2$ regularly alternate with the Heisenberg spins. The local anisotropy axis $z_1$ ($z_2$) of the Ising spins $\sigma=1/2$ on odd (even) lattice positions is canted by the angle $\alpha$ (-$\alpha$) from the global frame $z$-axis. Hence, it follows that the angle $2\alpha$ determines the overall canting between two coplanar local anisotropy axes $z_1$ and $z_2$. The Heisenberg spins $S=1/2$ are coupled to their nearest-neighbor Ising spins through the antiferromagnetic coupling $J<0$ projected into the respective anisotropy axis. Furthermore, we take into account the effect of the external magnetic field $B$, whose spatial orientation is given by the angle $\theta$ determining its tilting from the global frame $z$-axis. Under these circumstances, the spin-1/2 Ising-Heisenberg chain can be defined through the following Hamiltonian
\begin{align}
\mathcal{H} =& -J \sum_{i=1}^{N/2} (S_{2i-1}^{z_1} \sigma_{2i-1}^{z_{1}} \!+\! S_{2i-1}^{z_2} \sigma_{2i}^{z_{2}}
                                  \!+\! S_{2i}^{z_2} \sigma_{2i}^{z_{2}} \!+\! S_{2i}^{z_1} \sigma_{2i+1}^{z_{1}}) \nonumber\\
& -h^{z_{1}} \! \sum_{i=1}^{N/2}\! \sigma_{2i-1}^{z_{1}}
- h^{z_{2}} \! \sum_{i=1}^{N/2}\! \sigma_{2i}^{z_{2}} \nonumber\\
&- h^z \! \sum_{i=1}^{N}\! S_{i}^{z} - h^x \! \sum_{i=1}^{N}\! S_{i}^{x}\!,
\end{align}
where $h^{z_{1}} = g_{1}^{z_{1}} \mu_{\rm B} B $cos($\alpha-\theta)$ and $h^{z_{2}} = g_{1}^{z_{2}} \mu_{\rm B} B $cos($\alpha+\theta)$ determine  projections of the external magnetic field $B$ towards the anisotropy axes of the Ising spins on odd and even lattice positions, respectively,  $g_{1}^{z_{1}}$ and $g_{1}^{z_{2}}$ are the respective Land\'e g-factors of the Ising spins and $\mu_{\rm B}$ is the Bohr magneton. Similarly, $h^{z}= g_2^z \mu_{\rm B} B \cos \theta$ and $h^{x}= g_2^x \mu_{\rm B} B \sin \theta$ determine two orthogonal projections of the external magnetic field for the Heisenberg spins, whereas $g_2^z$ and $g_2^x$ are the respective spatial components of the Land\'e g-factors of the Heisenberg spins.

The total Hamiltonian of the spin-1/2 Ising-Heisenberg chain can be rewritten as the sum of the cell Hamiltonians
\begin{equation}
\mathcal{H}=\sum_{i=1}^{N/2}\left(\mathcal{H}_{2i-1}+\mathcal{H}_{2i}\right),\label{eq:2}
\end{equation}
each of which involves all the interaction and field terms of exactly one Heisenberg spin
\begin{align}
\mathcal{H}_{2i-1}=& -\frac{h^{z_{1}}}{2} \sigma_{2i-1}^{z_{1}} - \frac{h^{z_{2}}}{2} \sigma_{2i}^{z_{2}}
                     - h_{2i-1}^{z} S_{2i-1}^{z} - h_{2i-1}^{x} S_{2i-1}^{x}, \nonumber \\
\mathcal{H}_{2i}=& -\frac{h^{z_{2}}}{2} \sigma_{2i}^{z_{2}} - \frac{h^{z_{1}}}{2} \sigma_{2i+1}^{z_{1}} - h_{2i}^{z} S_{2i}^{z} - h_{2i}^{x} S_{2i}^{x}. \label{eq:3}
\end{align}
In above, we have introduced the following notation for the effective longitudinal and transverse fields acting on the Heisenberg spins
\begin{align}
  h_{2i-1}^{z} &=J\cos\alpha\left(\sigma_{2i-1}^{z_{1}}+\sigma_{2i}^{z_{2}}\right)+g_{2}^{z}\mu_{\rm B} B \cos \theta, \nonumber\\
  h_{2i}^{z}&=J\cos\alpha\left(\sigma_{2i}^{z_{2}}+\sigma_{2i+1}^{z_{1}}\right)+g_{2}^{z}\mu_{\rm B} B \cos \theta, \nonumber\\
  h_{2i-1}^{x}&=J\sin\alpha\left(\sigma_{2i-1}^{z_{1}}-\sigma_{2i}^{z_{2}}\right)+g_{2}^{x}\mu_{\rm B} B \sin \theta,\nonumber\\
  h_{2i}^{x}&=-J\sin\alpha\left(\sigma_{2i}^{z_{2}}-\sigma_{2i+1}^{z_{1}}\right)+g_{2}^{x}\mu_{\rm B} B \sin \theta. \label{eq:hef}
\end{align}
It is noteworthy that the cell Hamiltonians (\ref{eq:3}) commute and hence, they can be diagonalized independently of each other by performing a local spin-rotation transformation following the approach worked out previously \cite{str12}. In this way, one obtains the full spectrum of the eigenvalues, which can be subsequently utilized for the construction of the ground-state phase diagram and magnetization process. The full details of this calculation procedure will be published elsewhere together with a more comprehensive analysis of the thermodynamic properties.

\section{Results and discussion}

Let us illustrate a few typical ground-state phase diagrams and zero-temperature magnetization curves for the most interesting particular case with the antiferromagnetic coupling $J<0$, equal Land\'e g-factors of the Ising spins $g_1^{z_1}=g_1^{z_2}=20$ and equal components of the Land\'e g-factor of the Heisenberg spins $g_2^{x}=g_2^{z}=2$, which nearly coincide with usual values of gyromagnetic ratio for Dy$^{3+}$ and Cu$^{2+}$ magnetic ions, respectively. Under these circumstances, one finds four different ground states: two ground states CIF$_1$ and CIF$_2$ with the canted ferromagnetic alignment of the Ising spins
\begin{alignat}{1}
|{\rm CIF}_{1}\rangle= & \prod_{i=1}^{N/2}|\nearrow\rangle_{2i-1}|\psi\rangle_{2i-1}|\nwarrow\rangle_{2i}
|\psi\rangle_{2i},\label{eq:13}\\
|{\rm CIF}_{2}\rangle= & \prod_{i=1}^{N/2}|\swarrow\rangle_{2i-1}|\psi\rangle_{2i-1}
|\searrow\rangle_{2i}|\psi\rangle_{2i},\label{eq:14}
\end{alignat}
and two ground states CIA$_1$ and CIA$_2$ with the canted antiferromagnetic alignment of the Ising spins
\begin{alignat}{1}
|{\rm CIA}_{1}\rangle= & \prod_{i=1}^{N/2}|\nearrow\rangle_{2i-1}|\psi\rangle_{2i-1}
|\searrow\rangle_{2i}|\psi\rangle_{2i},\label{eq:15}\\
|{\rm CIA}_{2}\rangle= & \prod_{i=1}^{N/2}|\swarrow\rangle_{2i-1}|\psi\rangle_{2i-1}
|\nwarrow\rangle_{2i}|\psi\rangle_{2i}.\label{eq:16}
\end{alignat}
It is noteworthy that the state vector $|\nearrow\rangle_{2i-1}$ ($|\swarrow\rangle_{2i-1}$) corresponds to the spin state $\sigma_{2i-1}^{z_1} = 1/2$ ($\sigma_{2i-1}^{z_1} = -1/2$) of the odd-site Ising spins, the state vector $|\nwarrow\rangle_{2i}$ ($|\searrow\rangle_{2i}$) corresponds to the spin state $\sigma_{2i}^{z_2} = 1/2$ ($\sigma_{2i}^{z_2} = -1/2$) of the even-site Ising spins, while each Heisenberg spin
underlies a quantum superposition of both spin states
\begin{equation}
|\psi \rangle_i = \frac{1}{\sqrt{a_i^2 + 1}} \left(|\downarrow\rangle_i - a_i |\uparrow\rangle_i \right),
\label{eq:hv}
\end{equation}
which depends on the orientation of its two nearest-neighbor Ising spins via $a_i = h_i^x/[h_i^z - \sqrt{(h_i^z)^2 + (h_i^x)^2}]$.

\begin{figure}[htbp]
\vspace*{-1cm}
\includegraphics[width=0.49\columnwidth]{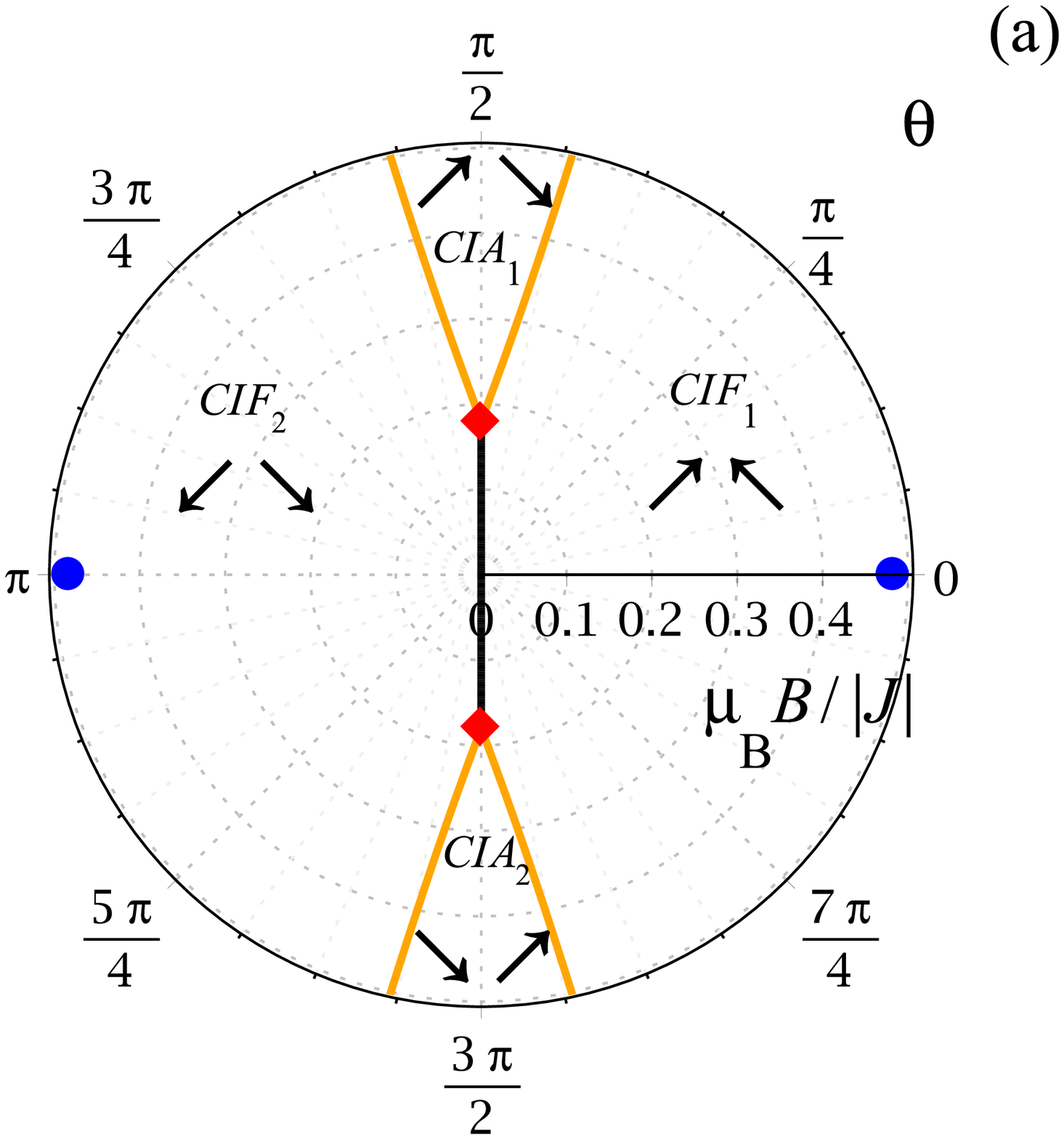}
\includegraphics[width=0.49\columnwidth]{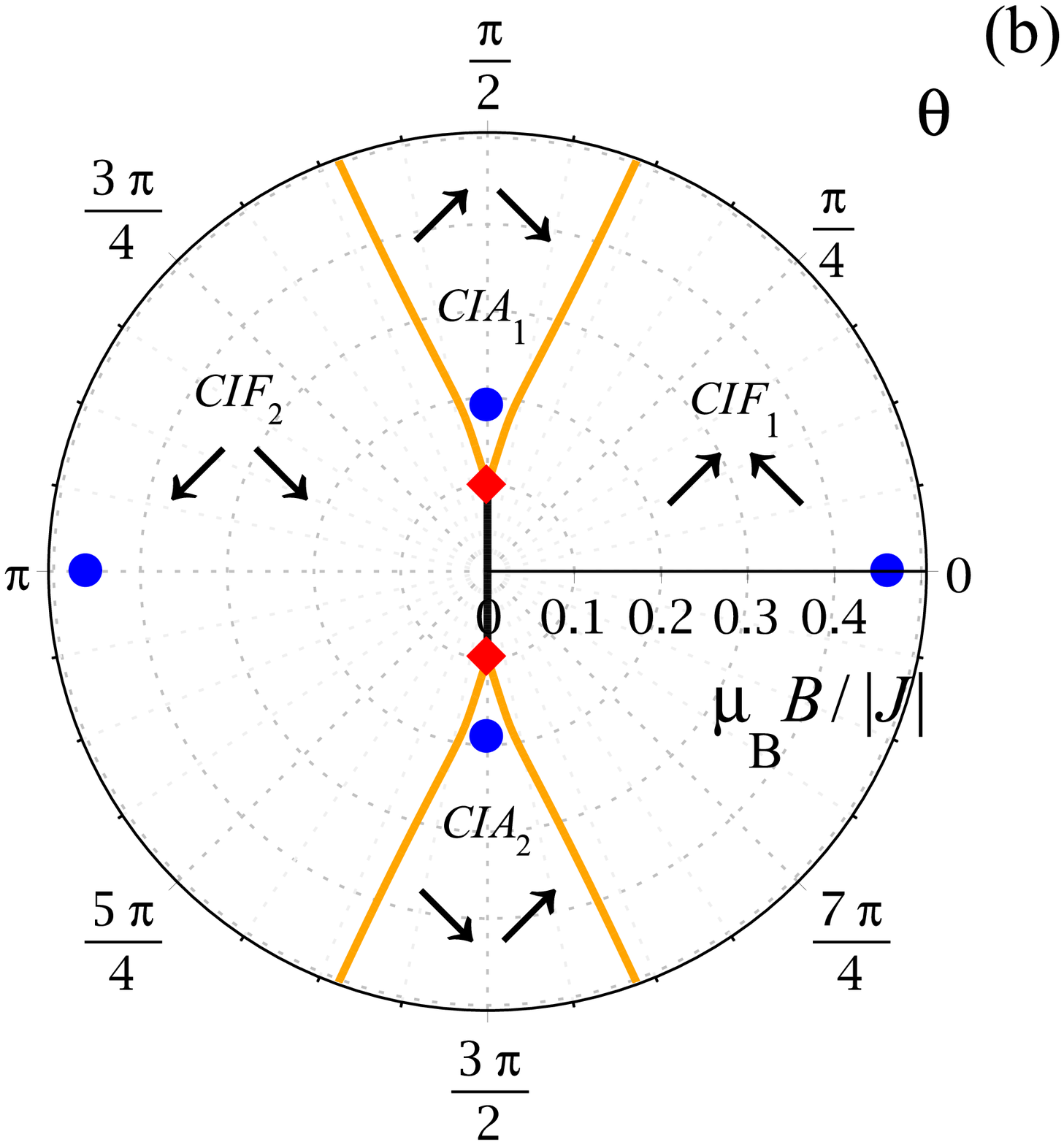}
\vspace*{0.2cm}
\caption{\label{fig:2} (Color online) The ground-state phase diagram in polar coordinates for two different canting angles between the local anisotropy axes:
(a) $2\alpha=\pi/6$; (b) $2\alpha=\pi/4$. The relative size of the magnetic field $\mu_{\rm B} B/|J|$ is represented by the radius of the polar coordinates and the angle $\theta$ determines its inclination with respect to the global frame $z$-axis.}
\end{figure}

The overall ground-state phase diagram in polar coordinates is illustrated in Fig.~\ref{fig:2} for two different canting angles $2\alpha$ between both local anisotropy axes. The phase diagram has an obvious symmetry with respect to $\theta = 0$ and $\pi/2$ axes, the former symmetry axis $\theta = 0$ merely interchanges CIA$_{1} \leftrightarrow$ CIA$_{2}$, while the latter symmetry axis $\theta = \pi/2$ is responsible for CIF$_{1} \leftrightarrow$ CIF$_{2}$ interchange. Without loss of generality, our further discussion will be therefore restricted just to the first quadrant $\theta \in [0, \pi/2]$. The coexistence line between CIF$_{1}$ and CIA$_{1}$ phases is macroscopically degenerate with the residual entropy per Ising-Heisenberg pair $S=k_{\rm B} \ln(2)/2$,  whereas CIF$_1$ and CIF$_2$ phases coexist together at $\theta=\pi/2$ up to a triple point (diamond symbol) with the residual entropy $S=k_{\rm B} \ln[(\sqrt{5}+3)/2]/2$. In addition, the Heisenberg spins are completely free to flip at macroscopically degenerate points (blue circles) with the residual entropy $S=k_{\rm B} \ln(2)$ given by the coordinates $B = |J|\cos(\alpha)/(2 \mu_{\rm B})$, $\theta=0$ and $B=|J|\sin(\alpha)/(2\mu_{\rm B})$, $\theta=\pi/2$ for $\alpha \gtrsim \pi/9$. These highly degenerate points correspond to a novel-type spin frustration 'half ice, half fire' \cite{yin16}, which originates from the difference between Land\'e g-factors being responsible for a fully frozen (ordered) character of the Ising spins and fully floppy (disordered) character of the Heisenberg spins.

\begin{figure}[t]
\includegraphics[width=\columnwidth]{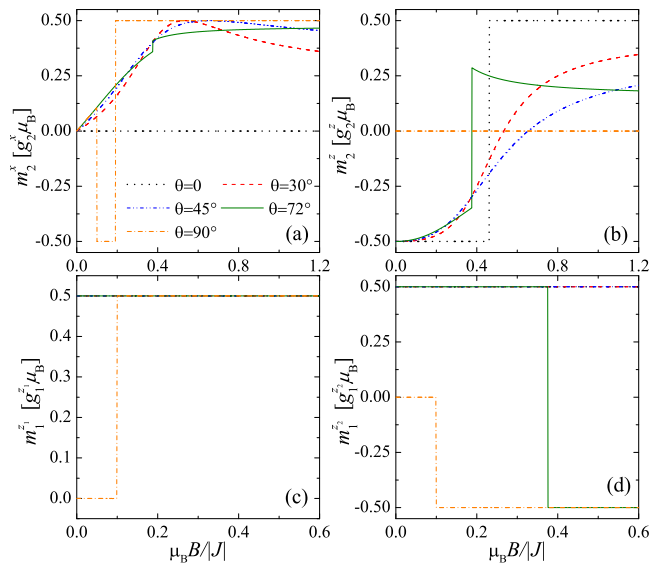}
\caption{\label{fig:3} (Color online) Zero-temperature magnetizations versus the relative strength of the magnetic field for the canting angle $2\alpha = \pi/4$ and several spatial orientations $\theta$ of the applied field: (a)-(b) the transverse $m_2^x$ and longitudinal $m_2^z$ projections of the Heisenberg spins; (c)-(d) the local projections $m_1^{z_{1}}$ and $m_1^{z_{2}}$ of the Ising spins towards their easy axes.}
\end{figure}

Finally, the individual contributions of the Ising and Heisenberg spins to the total magnetization are depicted in Fig.~\ref{fig:3} as a function of the magnetic-field strength for several spatial orientations $\theta$ of the applied field. As one can see, any deviation of the magnetic field from its longitudinal direction $\theta = 0$ destroys the sharp stepwise dependence in the longitudinal projection of the magnetization $m_2^z$ of the Heisenberg spins. In fact, the longitudinal component $m_2^z$ is gradually smeared out upon increasing of the tilting angle $\theta$, while its transverse part $m_2^x$ rises up to its global maximum successively followed by a gradual decline [cf. Fig.~\ref{fig:3}(a)-(b)]. The most notable dependences of the magnetization of the Heisenberg spins can be found for greater tilting angles $\theta$ of the magnetic field, which cause an abrupt jump in both components $m_2^x$ and $m_2^z$ of the Heisenberg spins due to a sudden reorientation of  Ising spins at the phase transition from the canted ferromagnetic phase CIF$_{1}$ to the canted antiferromagnetic phase CIA$_{1}$ (see the curves for $\theta = 2\pi/5$). The sharp stepwise dependence of the magnetization of the Heisenberg spins is afterwards recovered for the special case of the transverse field $\theta=\pi/2$, for which it appears in the transverse projection $m_2^x$ while its longitudinal part $m_2^z$ equals zero. The coexistence of the canted ferromagnetic phases CIF$_{1}$ and CIF$_{2}$ is manifested through a quasi-linear dependence of $m_2^x$ at low magnetic fields, where other contributions $m_2^z = m_1^{z_1} = m_1^{z_2} = 0$ vanish.

\section{Conclusion}
In the present work we have examined in detail the ground-state phase diagram and zero-temperature magnetization process of the spin-1/2 Ising-Heisenberg chain with two different local anisotropy axes in an arbitrarily oriented magnetic field. It has been shown that the phase diagram involves in total two canted ferromagnetic and two canted antiferromagnetic ground states. Another interesting finding concerns with the existence of a few macroscopically degenerate points, at which a perfect order of the Ising spins accompanies a complete disorder of the Heisenberg spins within the so-called 'half ice, half fire' frustrated ground state \cite{yin16}. It has been also convincingly evidenced that the canting angle between two local anisotropy axes of the Ising spins and the spatial orientation of the applied magnetic field basically influences the overall shape of the magnetization curves.

\section*{Acknowledgments}

This work was partially supported by FAPEAL (Alagoas State Research agency), CNPq, CAPES, FAPEMIG, VEGA 1/0043/16 and APVV-14-0073.


\begin{thebibliography}{99}

\bibitem{str05} J. Stre\v{c}ka, M. Ja\v{s}\v{c}ur, M. Hagiwara, K. Minami, Y. Narumi, K. Kindo,
\textit{Phys. Rev. B} \textbf{72}, 024459 (2005).
DOI:10.1103/PhysRevB.72.024459.

\bibitem{heu10} W. Van den Heuvel, L.F. Chibotaru,
\textit{Phys. Rev. B} \textbf{82}, 174436 (2010).
DOI:10.1103/PhysRevB.82.174436.

\bibitem{bel14} S. Bellucci, V. Ohanyan, O. Rojas,
\textit{EPL} \textbf{105}, 47012 (2014).
DOI: 10.1209/0295-5075/105/47012

\bibitem{sah12} S. Sahoo, J.P. Sutter, S. Ramasesha,
\textit{J. Stat. Phys.} \textbf{147}, 181 (2012).
DOI: 10.1007/s10955-012-0460-7.

\bibitem{str12} J. Stre\v{c}ka, M. Hagiwara, Y. Han, T. Kida, Z. Honda, M. Ikeda,
\textit{Condens. Matter Phys.} \textbf{15}, 43002 (2012).
DOI: 10.5488/CMP.15.43002.

\bibitem{cal08} G. Calvez, K. Bernot, O. Guillou et al.,
\textit{Inorg. Chim. Acta} \textbf{361}, 3997 (2008).
DOI: 10.1016/j.ica.2008.03.040.

\bibitem{yin16} W. Yin, Ch. Roth, A. Tsvelik, arxiv: 1510.00030v2.
\end{thebibliography}
\end{document}